\documentclass[st,twocolumn]{jpsj3}
\usepackage{txfonts}
\usepackage{graphicx}
\usepackage{longtable}
\usepackage{latexsym}
\usepackage{amsmath,amssymb}
\usepackage{color}
\usepackage{stackrel}
\usepackage{accents}
\usepackage{empheq}
\usepackage{cite}
\usepackage{xcolor}
\usepackage{wrapfig}

\title{Magneto-optical Spectroscopy with RAMBO: A Table-Top 30\,T Magnet}

\author{Fuyang Tay,$^{1,2}$\thanks{ft13@rice.edu} Andrey Baydin,$^{1,3}$\thanks{baydin@rice.edu} Fumiya Katsutani,$^{1}$ and Junichiro Kono$^{1,3,4,5}$}
\inst{$^1$Department of Electrical and Computer Engineering, Rice University, Houston, Texas 77005, USA\\
$^2$Applied Physics Graduate Program, Smalley-Curl Institute, Rice University, Houston, Texas, 77005, USA\\
$^3$Smalley-Curl Institute, Rice University, Houston, Texas, 77005, USA\\
$^4$Department of Physics and Astronomy, Rice University, Houston, Texas 77005, USA\\
$^5$Department of Materials Science and NanoEngineering, Rice University, Houston, Texas 77005, USA} 

\abst{
Optically probing materials in high magnetic fields can provide enlightening insight into field-modified electronic states and phases, while optically driving materials in high magnetic fields can induce novel nonequilibrium many-body dynamics of spin and charge carriers. While there are high-field magnets compatible with standard optical spectroscopy methods, they are generally bulky and have limited optical access, which prohibit performing state-of-the-art ultrafast and/or nonlinear optical experiments.  The Rice Advanced Magnet with Broadband Optics (RAMBO), a unique 30-T pulsed mini-coil magnet system with direct optical access, has enabled previously challenging experiments using femtosecond optical pulses, including time-domain terahertz spectroscopy, in cutting-edge materials placed in strong magnetic fields. Here, we review recent experimental advances made possible by the first-generation RAMBO setup. After summarizing technological aspects of combining optical spectroscopic techniques with the mini-coil magnet, we describe results of magneto-optical studies of a wide variety of materials, providing new insight into the states and dynamics of four types of quasiparticles in solids -- excitons, plasmons, magnons, and phonons -- in high magnetic fields. 
}

\begin{document}
\maketitle

\section{Introduction}
Cutting-edge materials such as topological insulators~\cite{HasanKane2010RMP}, low-dimensional materials~\cite{Goerbig2011RMP}, and strongly correlated materials~\cite{MorosanEtAl2012AM} generally show unusual properties in the presence of an external magnetic field $(B)$, which breaks the time-reversal symmetry of the system. Therefore, optical spectroscopy with access to a strong applied magnetic field is an indispensable tool for studying such materials. 

Numerous efforts have been made over the past decades to generate higher and higher magnetic fields for use in magneto-optical spectroscopy experiments, both in university and national laboratory settings. Currently, fields $B\lesssim$15\,T can be readily produced by commercial superconducting magnets with windows, whereas access to higher fields is typically limited to special facilities in national laboratories~\cite{HerlachMiura2003,CrowEtAl1996PBCM,SingletonEtAl2004PBCM,KiyoshiEtAl2006JPCS,WosnitzaEtAl2007JoMaMM,LiEtAl2010JLTP,DebrayFrings2013CRP}. Magnetic fields utilized for condensed matter research can be either static or pulsed. Pulsed magnets can generally produce higher magnetic fields~\cite{MiuraHerlach1985Book} and even peak fields $B>100$\,T can be generated for magneto-optical experiments~\cite{KonoetAl93Physica,KonoMiura06HMF,BooshehrietAl12PRB,JaimeEtAl2012P,WangEtAl2020ITAS,NakamuraEtAl2018RSI}. 

Although some of these magnet systems are designed with optical access, there are limitations on the types of optical measurements that can be performed in high magnetic fields. The primary issue is that optical access is usually available through optical fibers due to the large size of the magnet systems. This raises challenges in optical alignment, and thus, the optical setup cannot be easily modified for different types of optical measurements, e.g., photoluminescence (PL) spectroscopy, magneto-optic Kerr effect measurements, and terahertz time-domain spectroscopy (THz-TDS). The dispersion and/or pulse broadening that occurs for light propagating in the optical fibers severely restricts possible ultrafast optical measurements, polarization-sensitive measurements, and experiments involving broadband radiation. Additionally, in pulsed high-field magnets, it takes the magnet a long time to cool down after each shot, which results in the wait time between successive magnetic shots ranging from several minutes to more than an hour.

Here, we review magneto-optical studies performed with the Rice Advanced Magnet with Broadband Optics (RAMBO), a unique table-top, mini-coil pulsed magnet system~\cite{NoeEtAl2013RSI}. The RAMBO magnet can generate a magnetic field pulse with a peak field up to 30\,T, and the sample cryostat provides temperature control between 12\,K and 300\,K. RAMBO has direct optical access through optical windows, which allows ultrafast optical spectroscopy experiments with minimal pulse broadening. The portability and small footprint of RAMBO permit its convenient incorporation into any optical spectroscopy setup. 

This review article is organized as follows. First, in Section~\ref{sec:exp}, we introduce the configuration of RAMBO and several optical spectroscopies that have been used with it. Next, in Section~\ref{sec:material}, we discuss representative magneto-optical studies of various excitations in materials such as excitons, plasmons, magnons, and phonons that have been successfully performed with the RAMBO system. Finally, in Section~4, we conclude this review with an outlook for the future of RAMBO and high-field magneto-optical studies of condensed matter. 


\section{Experimental Methods}
\label{sec:exp}
\subsection{The magnet system}
\begin{figure}[h]
    \centering
    \includegraphics[width=0.45\textwidth]{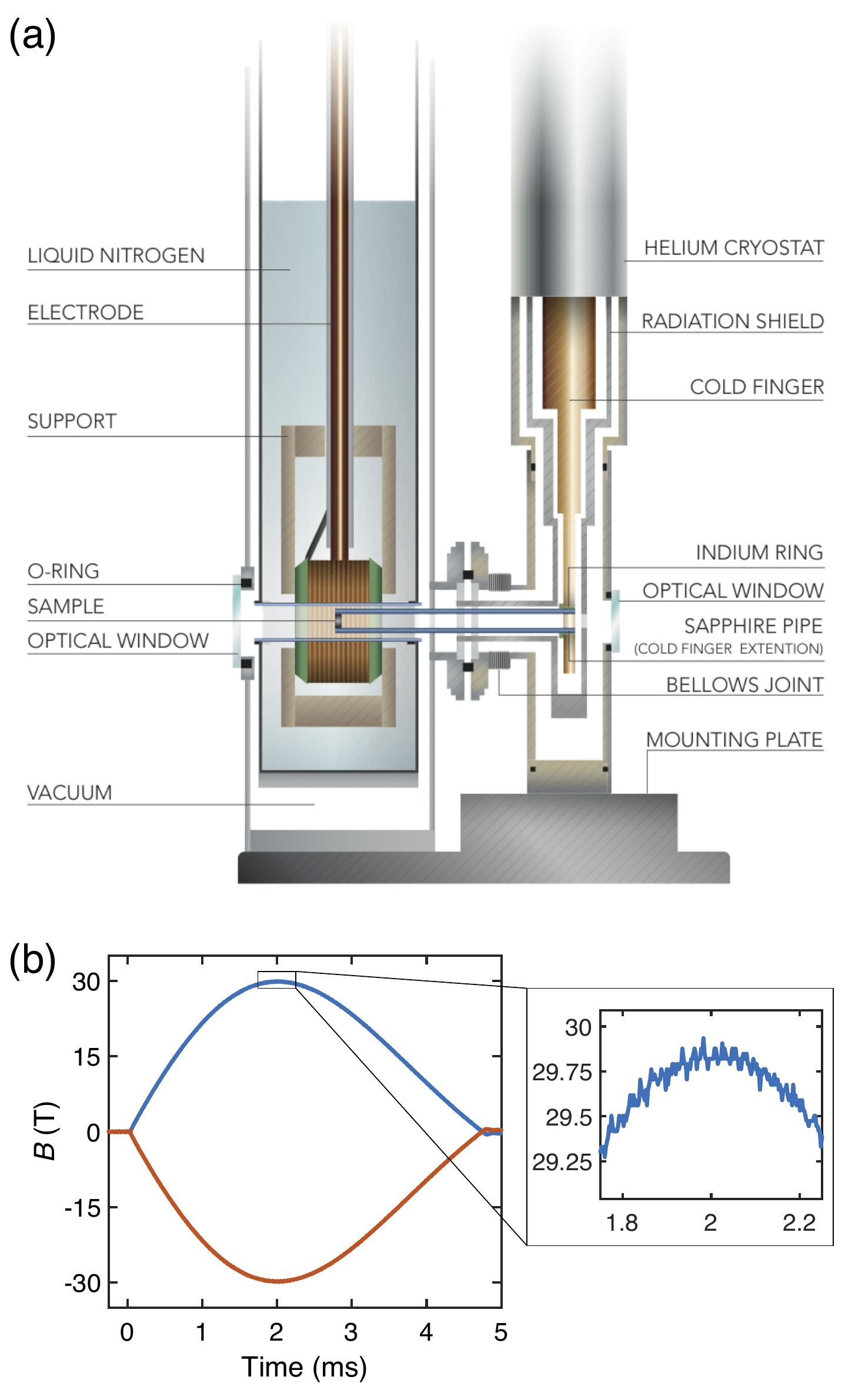}
    \caption{(a)~Schematic diagram of the RAMBO system. Illustration by Tanyia Johnson. (b)~Magnetic field pulse time profile with a peak field of $+30$\,T (blue) and $-30$\,T (red). The inset displays an expanded view of the magnetic field profile around the peak, showing a magnetic field variation of $\sim0.5$\,T (or $\sim2$\,\%) in a time duration of $\sim0.5$\,ms.}
    \label{fig:RAMBO}
\end{figure}

The RAMBO system consists of two cryostats -- one cryostat for keeping the magnet cold, and one cryostat for controlling the temperature of the sample under study. The magnet cryostat (shown on the left in Fig.~\ref{fig:RAMBO}a), which contains the magnet coil, is always filled with liquid nitrogen to ensure that the magnet is cold before each shot as well as to rapidly cool down the magnet after each shot. The sample cryostat (shown on the right in Fig.~\ref{fig:RAMBO}a) is a commercial liquid helium flow cryostat (Cryo Industries, Inc., CFM-1738-102).

The sample is attached to the tip of a tapered sapphire pipe in such a way that it is located at the center of the magnet bore. The bottom of the sapphire pipe was attached to the cryostat cold finger. The lowest achievable temperature is $\sim$12\,K, measured through a temperature sensor attached to the tip of the sapphire pipe close to the sample temperature. A heater is mounted on the cold finger to vary the sample temperature between 12\,K and 300\,K. The two cryostats are connected with a vacuum pipe, and they share the same vacuum space.  Optical windows are held on both ends of the magnet system, enabling optical experiments in a transmission geometry.

Two coils have been constructed for RAMBO. Detailed information on the first coil is provided in a previous publication~\cite{NoeEtAl2013RSI}. The second-generation magnet coil is made by winding a 1\,mm\,$\times$\,1.5\,mm AgCu wire around the metal bore. The coil is composed of 13 layers, and each layer consists of 14-15 turns. The total wire length is approximately 16.5\,m. The inductance of the coil at 77\,K is 411.7\,$\mu$H, which is close to the previous coil~\cite{NoeEtAl2013RSI}. 

A pickup coil is wrapped around the sapphire pipe near the sample to measure the generated magnetic field pulse. Figure~\ref{fig:RAMBO}b shows a measured temporal profile of a magnetic field pulse with a peak field of $\pm$30\,T and a pulse duration of about 4.7\,ms. The magnetic field reaches its maximum at about 2\,ms after the capacitor bank discharge process is initiated by the trigger pulse. The inset in Fig.~\ref{fig:RAMBO}b shows that the magnetic field value only changes by about 2\,\% within 500\,$\mu$s around the peak. Therefore, it is reasonable to assume that the sample experiences a constant magnetic field at the peak value for the duration of 500\,$\mu$s. A versatile magnet control unit is used to synchronize the magnetic field pulse with other instruments such as the output from the amplified Ti:sapphire laser.

The maximum sample size is restricted by the diameter of the magnet bore, which is 12\,mm. The numerical aperture of the sapphire pipe is 0.03~\cite{NoeEtAl2013RSI}. Nevertheless, the cylindrical sapphire pipe has been replaced by a conical sapphire pipe with an outer diameter of 8\,mm at the side to mount the sample and an outer diameter of 12.6\,mm at the side contacted to the cold finger, in order to have an even larger numerical aperture.

\subsection{Time-resolved photoluminescence spectroscopy}
\begin{figure}[h]
    \centering
    \includegraphics[width=0.5\textwidth]{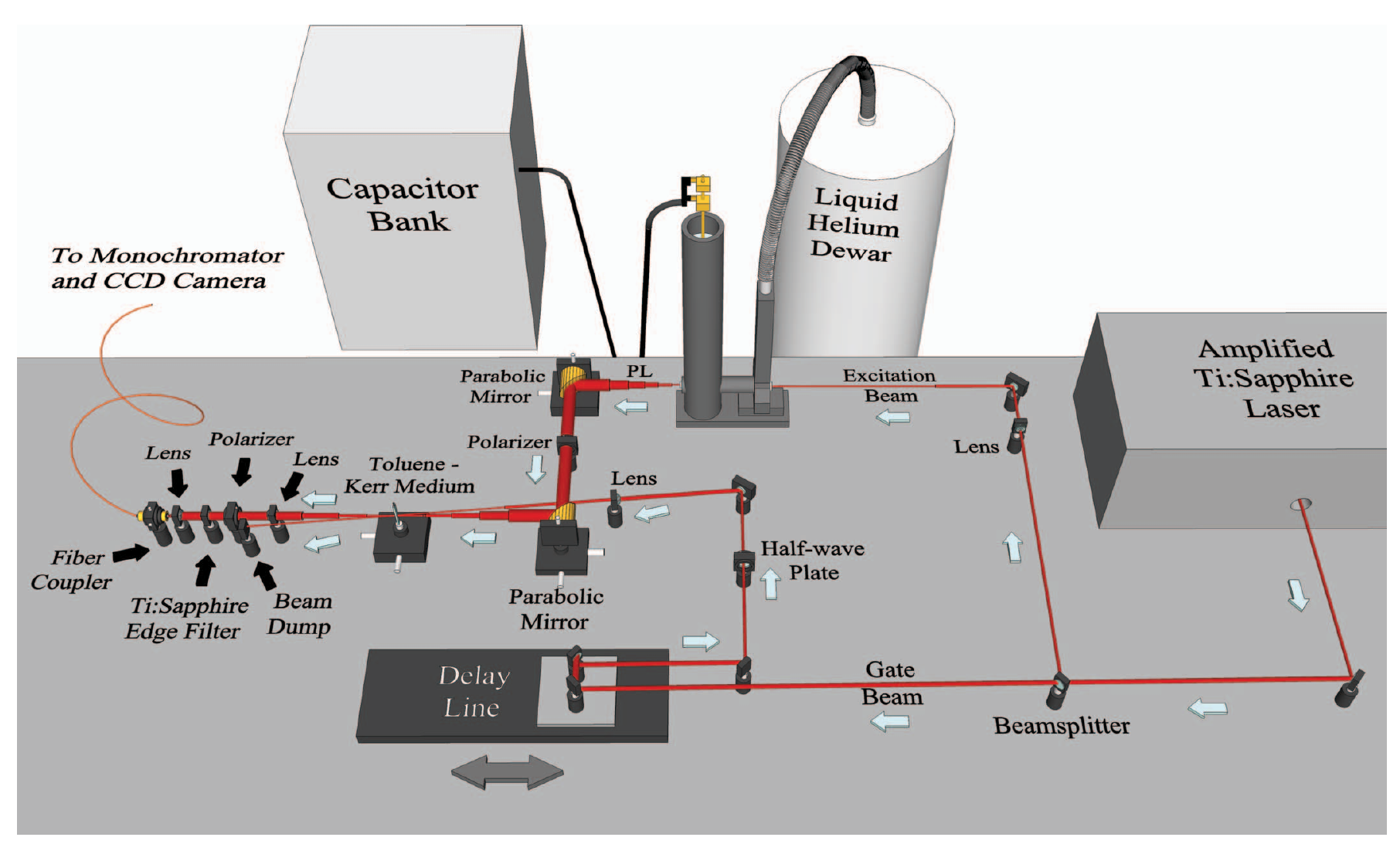}
    \caption{Illustration of the time-resolved photoluminescance spectroscopy system using the optical Kerr gating method. The output from the amplified Ti:sapphire laser is split into an excitation beam and a gate beam. The light blue arrows indicate the propagation direction of the excitation beam. After excitation, the PL emitted by the sample is collected and focused onto a Kerr medium placed between a pair of crossed polarizers. The gate beam induces transient birefringence in the Kerr medium so that the PL that passes through the medium simultaneously is rotated and can then partially pass through the second polarizer. A delay line is used to change the time delay between the excitation beam and the gate beam to map out the emission dynamics. Adapted from Ref.~\citenum{NoeEtAl2013RSI}.}
    \label{fig:timeresolvedPL}
\end{figure}

The RAMBO system has been incorporated into a time-resolved PL spectroscopy setup, which is shown in Fig.~\ref{fig:timeresolvedPL}, to measure ultrafast light emission dynamics at high magnetic fields~\cite{NoeEtAl2013RSI}. The output of the Ti:sapphire regenerative amplifier (Clark-MXR, Inc., CPA-2001) centered at 775\,nm with 150\,fs pulse duration, 1\,kHz repetition rate, and pulse energy up to 5\,$\mu$J is split into an excitation beam and a gate beam. The excitation beam is focused onto the sample inside the cryostat with a spot size of about 500\,$\mu$m. A chopper is used to reduce the repetition rate of the excitation beam to 50\,Hz. The magnet control unit is synchronized with the pulsed laser so that the excitation pulse hits the sample at the peak of the magnetic field pulse. As discussed before, the magnetic field variation is negligible for transient phenomena of the order of picoseconds excited by a femtosecond pulse.

Two off-axis parabolic mirrors are used to collect and refocus the emission onto a Kerr medium, toluene. Two crossed polarizers are placed before and after the toluene. Without a gate pulse, the PL is blocked by the crossed polarizers. Conversely, the Kerr medium rotates the polarization of the PL when a gate pulse with a polarization of 45$^\circ$ with respect to the crossed polarizers hits the Kerr medium simultaneously, allowing some of the PL to pass the second polarizer into the detector scheme. The path distance of the gate pulse is controlled by a delay line to map out the PL dynamics. After the Kerr medium, a long-pass filter is used to eliminate the gate pulse. The PL is focused onto an optical fiber and measured with a silicon charge-coupled device (CCD) camera attached to a grating spectrometer.  

\subsection{Time-integrated photoluminescence spectroscopy}
The time-integrated PL spectroscopy setup is the same as the time-resolved PL setup depicted in Fig.~\ref{fig:timeresolvedPL} except that the crossed polarizers, the Kerr medium (toluene), and the gate beam are removed. In addition to the strong pulsed excitation source, a laser diode (World Star Tech, TECiRL-15G-780) and a continuous-wave Ar ion laser (Melles Griot) are used as light sources for different measurements. Both light sources are modulated to be turned on for a short duration ($<$500\,$\mu$s) at the peak of the magnetic pulse.

\subsection{Transmission spectroscopy}
Two light sources are used for transmission measurements, a light-emitting diode (LED) (Thorlabs, Inc., LED880L-50) and a supercontinuum laser (EXU6, NKT Photonics). The supercontinuum laser generates pulses with a repetition rate of 80\,MHz, a pulse duration of $>$1\,ps, and a pulse energy of around 0.01\,nJ. Light transmits through the sample inside the cryostat, and a spectrometer is used to detect and analyze the transmitted light. The light beams are modulated to study the transmission of the sample only at the peak magnetic field. A quarter-wave plate is used to create circularly polarized light for some measurements.

\begin{figure}[h]
    \centering
    \includegraphics[width=0.48\textwidth]{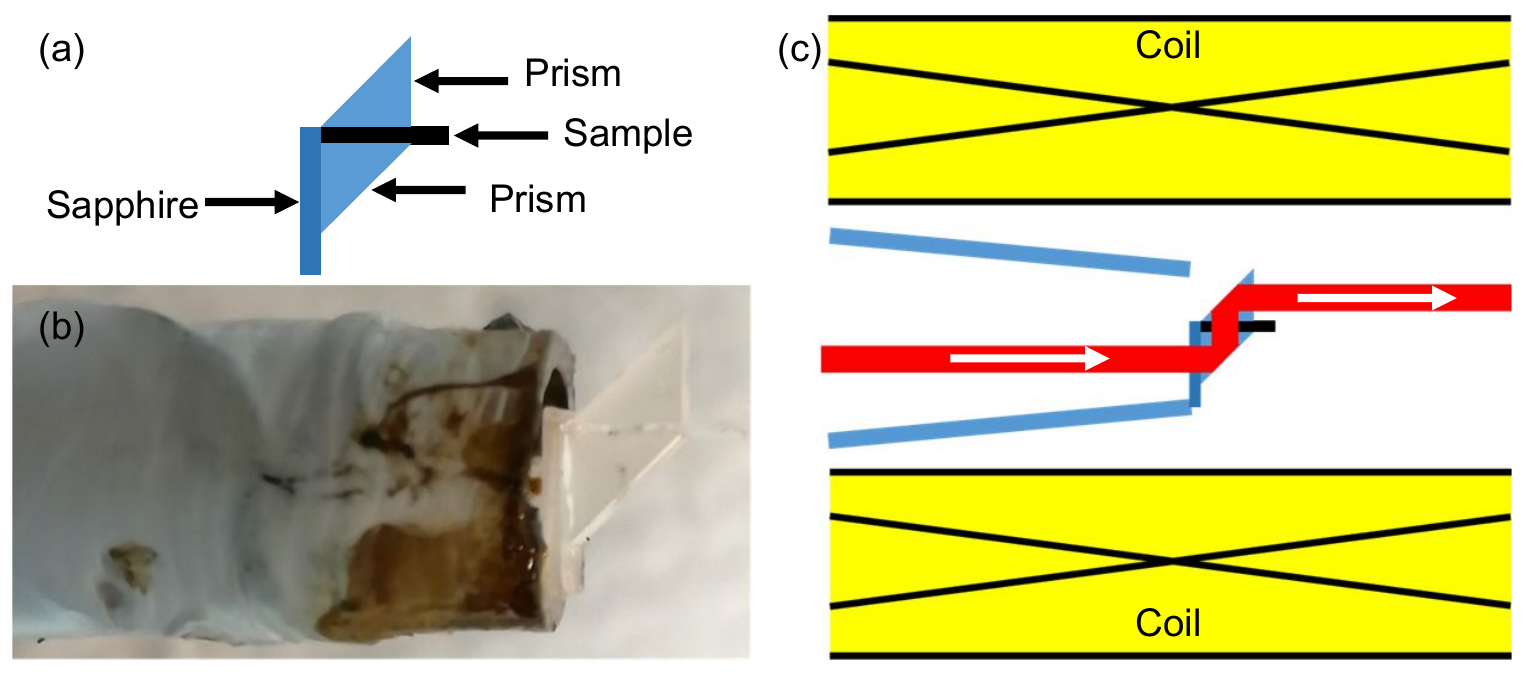}
    \caption{(a)~Illustration of the sample holder for transmission measurements in the Voigt configuration. The sample is sandwiched between a pair of prisms and attached to a sapphire substrate. (b)~Image of the fabricated sample. (c)~Schematic diagram of the Voigt configuration setup. The blue solid lines represent the tapered sapphire pipe, and the red solid line indicates the light propagation path.}
    \label{fig:voigt}
\end{figure}

Measurements in the Voigt geometry can also be made with the RAMBO system. A pair of prisms can be used to sandwich the sample, as shown in Fig.~\ref{fig:voigt}a. The light beam passes through the sample along the normal direction after the first total internal reflection, and then it is reflected back by the second prism to leave the cryostat. The magnetic field is parallel to the sample surface and perpendicular to the light propagation direction.

\subsection{Single-shot terahertz time-domain spectroscopy}
A broadband terahertz (THz) pulse (0.25-1.6\,THz) is generated either in a Mg-doped stoichiometric LiNbO$_3$ crystal with the tilted-pulse-front excitation method~\cite{HeblingEtAl2008JOSABJ} or in a ZnTe crystal by optical rectification for a broader bandwidth. The emitter crystal is pumped by the output of an amplified Ti:sapphire laser (Clark-MXR, Inc., CPA-2001) centered at 775\,nm with 150\,fs pulse duration and 1\,kHz repetition rate. The generated THz pulse is focused onto the sample mounted on the sapphire pipe with a pair of parabolic mirrors. 

The transmitted THz radiation is detected through electro-optic sampling~\cite{Lee2009,PrasankumarTaylor2012} with a (110) ZnTe detection crystal. In conventional THz-TDS, a THz waveform is recorded in a point-by-point manner with the step-scan method~\cite{MolterEtAl2012OEO}, where the measurement is repeated for different time delays between the THz pulse and the gate pulse arriving at the detection crystal. In this way, only one data point of the THz waveform can be collected in a magnet shot. However, the wait time between shots of our magnet system is on the order of several minutes for shots up to 30\,T. Therefore, this method is time-consuming and inefficient to obtain THz waveforms. 

\begin{figure}[h]
    \centering
    \includegraphics[width=0.48\textwidth]{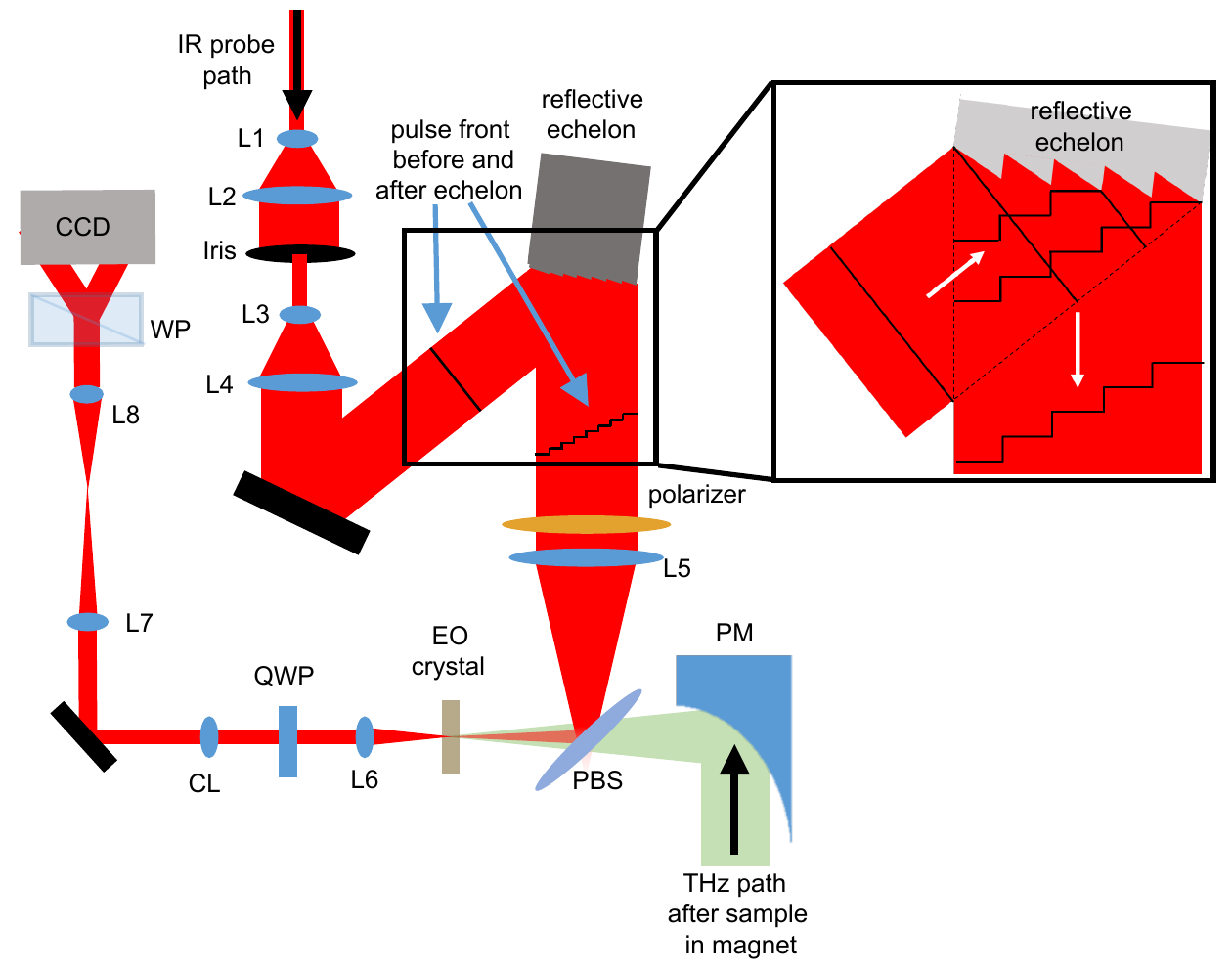}
    \caption{Schematic diagram of the detection scheme of the single-shot THz-TDS. L1-L4 are lenses to expand the infrared gate beam so that the transmitted gate beam has a relatively uniform intensity profile. Time delay information is encoded onto the intensity profile of the gate beam upon reflection from the echelon mirror. The gate beam overlaps with the THz beam transmitted through the sample by using a pellicle beamsplitter, PBS. The beams are focused onto an electro-optic sampling crystal, EO crystal. A  quarter-wave plate, QWP, a cylindrical lens, CL, and a Wollaston prism, WP, are used to detect the birefringence in the EO crystal induced by the electric field of the THz beam. Adapted from Ref.~\citenum{NoeEtAl2016OEO}.}
    \label{fig:singleshotTHz}
\end{figure}

Different rapid scanning methods have been demonstrated to overcome this issue~\cite{MolterEtAl2010OE,NoeEtAl2014AOA,SpencerEtAl2016APL,BaydinEtAl2021FO}. We employ the single-shot THz detection technique in our THz-TDS setup~\cite{YellampalleEtAl2007OEO,MinamiEtAl2013APL,TeoEtAl2015RSI,NoeEtAl2016OEO}. A large reflective echelon mirror is utilized to encode time delay information on a single gate pulse~\cite{KatayamaEtAl2011JJAP,MinamiEtAl2013APL,MinamiEtAl2015APL,NoeEtAl2016OEO,MeadEtAl2019RSI}. The size of the customized echelon mirror is 20\,mm\,$\times$\,20\,mm, and it consists of 1000 steps of 20\,$\mu$m width and a step height of 5\,$\mu$m. As shown in the inset of Fig.~\ref{fig:singleshotTHz}, the reflected gate beam exhibits a stair-step wavefront profile with the step height corresponding to an incremental delay of $\sim$33\,fs. The initial 150\,fs gate pulse becomes a $\sim$33\,ps pulse upon reflection from the echelon mirror. Two 10$\times$ telescopes are used to expand the gate beam before the echelon mirror so that the intensity profile of the beam that reflects off of the echelon mirror is relatively uniform. A quarter-wave plate, a Wollaston prism, and a silicon CCD camera are placed at the path of the gate pulse after the detection crystal to measure the induced birefringence in the crystal that is proportional to the electric field amplitude of the THz pulse~\cite{Lee2009,PrasankumarTaylor2012}. 

A pair of wire-grid polarizers is placed before and after the sample to ensure that the THz pulse is linearly polarized. The second polarizer after the sample can be rotated by 90$^\circ$ for Faraday and Kerr rotation measurements. It should be noted that both Faraday and Kerr rotation signals can be extracted in a transmission configuration~\cite{ValdesAguilarEtAl2012PRL,ShimanoEtAl2013NC,WuEtAl2016S,LiEtAl2019PRB}. The THz pulse that directly transmits through the sample contains only the Faraday rotation signal. Nevertheless, the back reflection pulse experiences both Faraday and Kerr rotation due to additional reflection events at the interfaces. Therefore, the Kerr rotation signal can be extracted by subtracting the Faraday rotation from the polarization rotation of the back reflection pulse. 

For optical-pump/THz-probe spectroscopy experiments, an additional beam splitter is used to split the near-infrared pulse into the pump beam for sample excitation and the beam for THz generation and detection~\cite{NoeEtAl2016OEO}. The optical delay between the optical pump and the THz probe pulses is controlled by a linear delay stage.

\section{Magneto-optical Studies of Excitons, Plasmons, Magnons, and Phonons with RAMBO}
\label{sec:material}
\subsection{Excitons}
An exciton is a bound state of an electron in the conduction band and a hole in the valence band due to the Coulomb interaction. It usually forms when a material is excited by a photon of energy higher than the band gap. The envelope wavefunction of the exciton is analogous to that of a hydrogen atom. This section only considers Wannier-Mott excitons in semiconductors. Owing to the large dielectric constant in semiconductors and smaller reduced mass of the exciton, the Wannier-Mott exciton has a Bohr radius much larger than the lattice spacing and a small binding energy ($\sim$1-100\,meV).

\begin{figure}[h]
    \centering
    \includegraphics[width=0.48\textwidth]{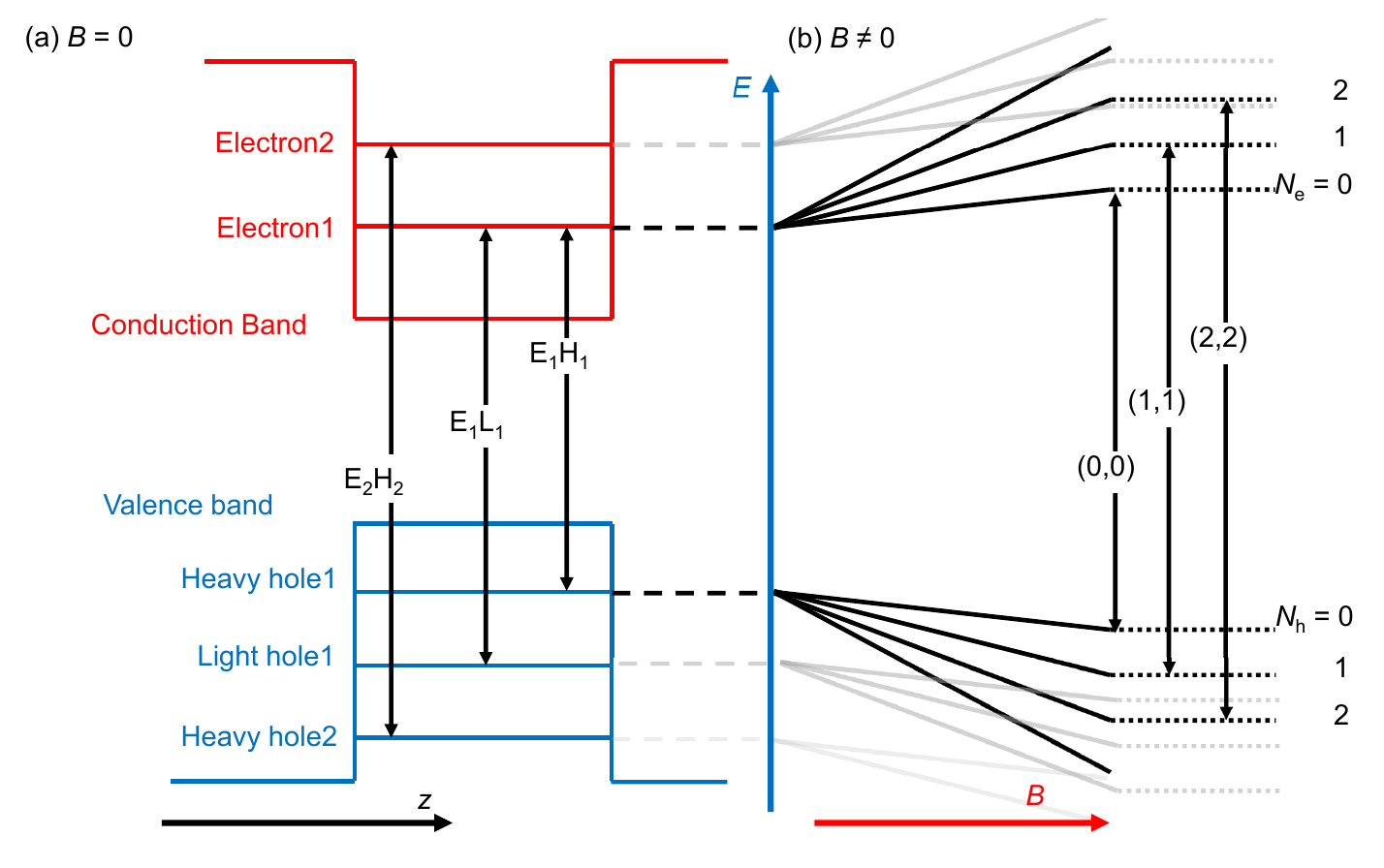}
    \caption{(a)~Illustration of the electron energy states for a QW at $B=0$. Two electron subbands and three hole subbands are considered in this case. (b)~The electron and hole subbands split into multiple Landau levels (LL) when a magnetic field is applied perpendicular to the QW.}
    \label{fig:energylevels}
\end{figure}

An example of the energy states for a quantum well (QW) without considering Coulomb interactions is illustrated in Fig.~\ref{fig:energylevels}. Two electron subbands and three hole subbands are confined in the QW. A notation is used to denote the transition between two bands, for instance, E$_1$H$_1$ represents the transition between the electron\,1 band and the heavy hole\,1 band. When an external magnetic field is applied along the growth direction, the states split due to Landau quantization. $N_\text{e}$ and $N_\text{h}$ denote the electron and hole Landau quantum numbers, respectively. 

The magnetic field, $B$, dependence of the energies of excitonic states is critically dependent on key exciton parameters such as the reduced effective mass, $\mu^*$, binding energy, effective Bohr radius, and effective $g$-factor, $g^*$. Therefore, magneto-optical spectroscopy has been commonly used to determine these important materials parameters for various semiconductors.  

Approximate solutions for excitonic levels can be obtained in the low-$B$ limit or the high-$B$ limit. A dimensionless parameter, $\gamma=16\pi^2\hbar^3\varepsilon^2B/\mu^{*2}e^3$~\cite{CongEtAl2018EoMO}, is introduced to define the low-$B$ and high-$B$ regimes where $\hbar$ is the reduced planck constant, $\varepsilon$ is the permittivity, and $e$ is the electronic charge. In the low-$B$ regime ($\gamma\ll1$), the magnetoexciton transition energy, $E$, can be approximately expressed as~\cite{Miura2008,CongEtAl2018EoMO}
\begin{equation}
\label{eqn:lowB}
    E(B) = E(B=0)+\sigma B^2 \pm g^*\mu_\text{B}B,
\end{equation}
where $\mu_\text{B}$ is the Bohr magneton and $\sigma$ is the diamagnetic-shift coefficient. The second term is the diamagnetic term that is proportional to $B^2$ and the last term is the energy splitting term due to the Zeeman effect, which can be probed by circularly polarized light. On the other hand, in the high-$B$ regime ($\gamma\gg1$), $E$ exhibits a linear $B$ dependence and it can be fit by the equation below~\cite{Miura2008,CongEtAl2018EoMO}
\begin{equation}
\label{eqn:highB}
    E(B) = E_g+\left(N+\frac{1}{2}\right)\frac{\hbar eB}{\mu^*}\pm g^*\mu_\text{B}B,
\end{equation}
where $E_g$ and $N$ are the band gap and Landau quantum number, respectively. 

\subsubsection{Magnetoexcitons in InGaAs/GaAs quantum wells}
\begin{figure}[h]
    \centering
    \includegraphics[width=0.5\textwidth]{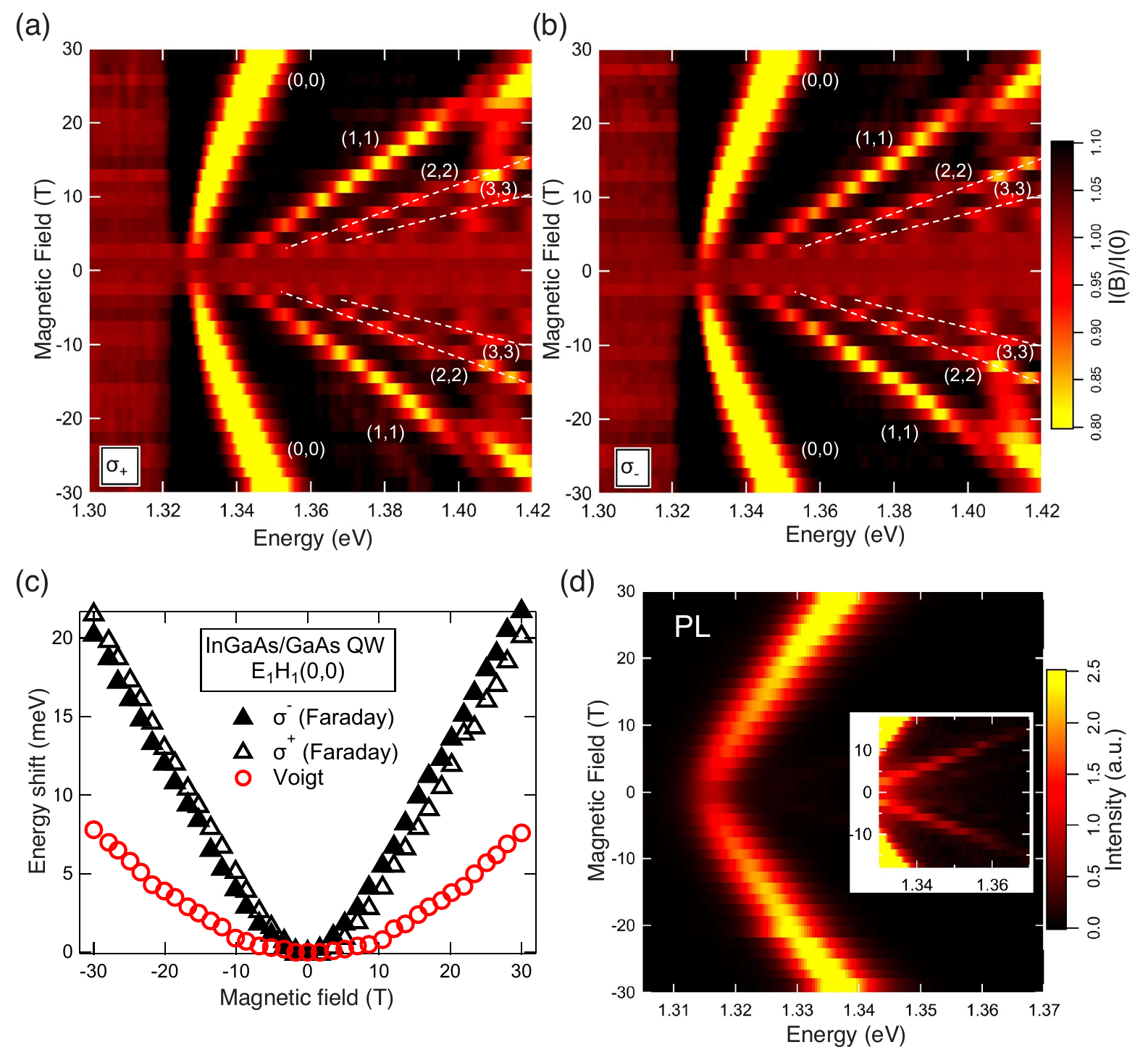}
    \caption{Magnetic-field-dependent $I(B)/I(0)$ spectral map probed by (a)~$\sigma_+$ and (b)~$\sigma_-$ polarizations, respectively. (c)~Energy shifts of the E$_1$H$_1$\,(0,0) peak measured in the Faraday geometry using probe beams with $\sigma_+$ and $\sigma_-$ polarization and measured in the Voigt geometry. (d)~Magneto-PL spectrum from $-30$\,T to $30$\,T. The inset displays an expanded view of the PL map.}
    \label{fig:InGaAs}
\end{figure}

The InGaAs/GaAs sample was grown by molecular beam epitaxy and contained 15 periods of 15-nm-wide InGaAs QWs with 8-nm-wide GaAs barriers. Figures~\ref{fig:InGaAs}a and \ref{fig:InGaAs}b show the ratios of the intensity of transmitted light at a finite magnetic field, $I(B)$, to the intensity of transmitted light at zero magnetic field, $I(0)$, using light with different circular polarizations ($\sigma_+$ and $\sigma_-$). The supercontinuum laser was used as the light source. The magnetic field dependence of magnetoexcitonic transition energies from E$_1$H$_1$\,(0,0) to E$_1$H$_1$\,(3,3) are mapped out, and $\mu^*$ is calculated to be (0.065\,$\pm$\,0.05)$m_0$ according to Eq.~(\ref{eqn:highB}).

The sample was also measured in the Voigt geometry where the applied magnetic field was parallel to the quantum well plane. The energy shifts of the E$_1$H$_1$(0,0) transition from the Faraday and Voigt configuration measurements are plotted in Fig.~\ref{fig:InGaAs}c. The energy shift in the Voigt geometry is smaller than that in the Faraday geometry because the size of the exciton perpendicular to the QW is smaller.

Figure~\ref{fig:InGaAs}d displays the result of magneto-PL measurements using the Ar ion laser at a wavelength of 488\,nm. The E$_1$H$_1$\,(0,0) exciton peak at 0\,T in the PL measurement is 6\,meV lower than that observed in the absorption spectrum, which is likely due to alloy disorder in the QW. The PL intensity and energy increase as $B$ increases. The emission from the E$_1$H$_1$\,(1,1) state is shown in the inset.

\subsubsection{Magnetoexcitons in InSe}
\begin{figure}[h]
    \centering
    \includegraphics[width=0.45\textwidth]{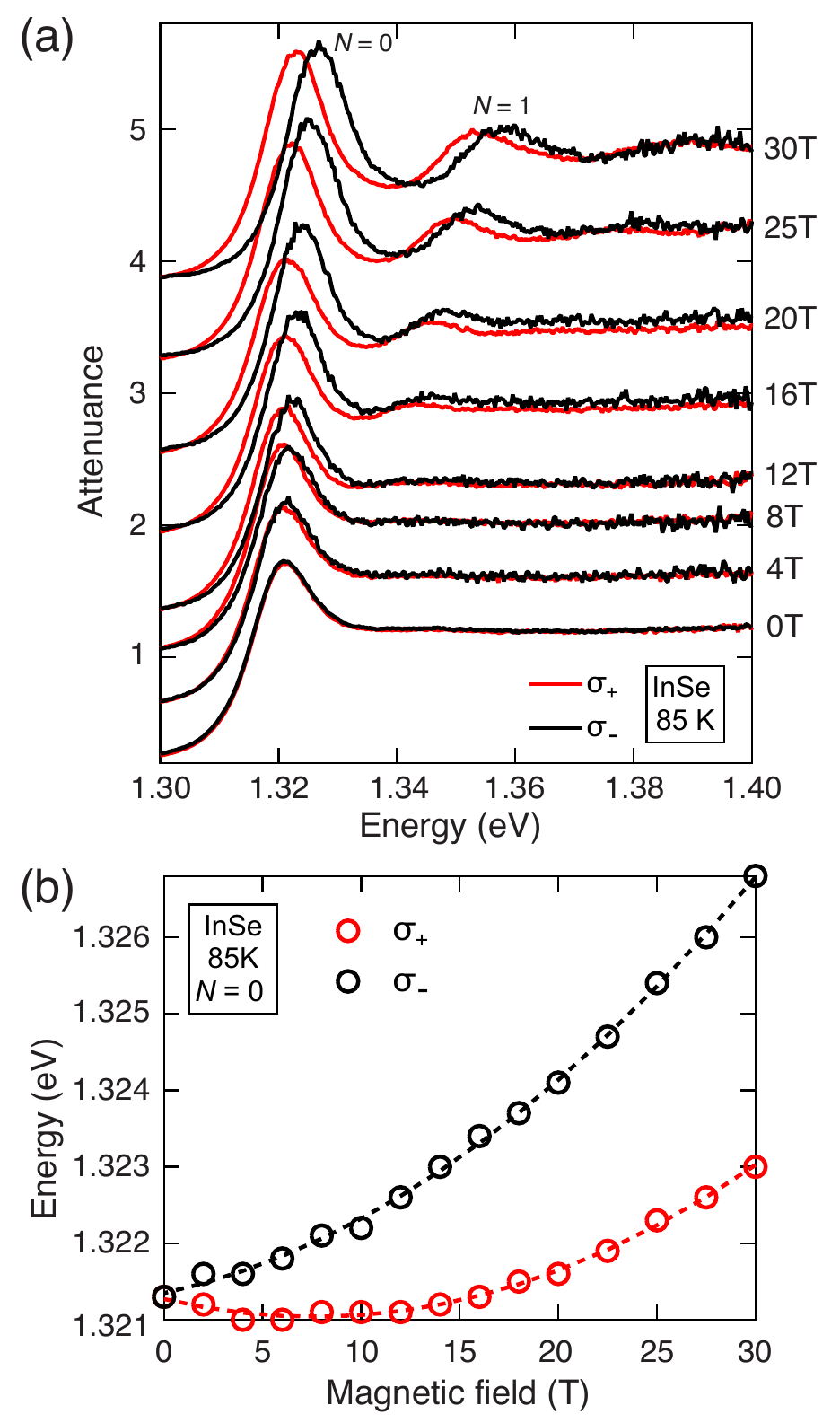}
    \caption{(a)~Attenuation spectra of InSe  up to 30\,T at 85\,K (b)~The peak energy of $N=0$ transitions extracted from (a). The dashed lines are fits to the data.}
    \label{fig:InSe}
\end{figure}

A $\sim$50\,$\mu$m thick InSe thin film was prepared by cleaving a single crystal. The magnetic field was applied along the $c$-axis of InSe. Figure~\ref{fig:InSe}a shows magnetoabsorption spectra up to 30\,T probed by the $\sigma_+$ and $\sigma_-$ polarizations at 85\,K. The lowest exciton peak is observed at 1.321\,eV at 0\,T. The peak shifts with increasing magnetic field and a second peak emerges at high $B$. The peaks at $B\neq0$ are attributed to $N=0$ and $N=1$ transition. The $N=0$ peak energies are plotted in Fig.~\ref{fig:InSe}c and fit by Eq.~(\ref{eqn:lowB}). The $g^*$ and $\sigma$ are obtained as $1.06\pm0.04$ and $4.08\times10^{-3}$\,meV/T$^2$, respectively.

\subsubsection{Superfluorescence from In$_{0.2}$Ga$_{0.8}$As quantum wells}
When numerous two-level dipoles are initially prepared in an incoherent state and confined in a small volume, they spontaneously develop macroscopic coherence from vacuum fluctuations, leading to a burst of radiation after a finite delay time. This phenomenon is known as superfluorescence (SF),~\cite{Dicke1954PRa,BonifacioLugiato1975PRA,BonifacioLugiato1975PRAa,VrehenEtAl1980N,SchuurmansEtAl1982AiAaMP} and it was first observed in atomic and molecular systems~\cite{SkribanowitzetAl73PRL,GrossetAl76PRL,GibbsetAl77PRL}. The observations of SF in semiconductors had been challenging owing to the fast scattering rates of carriers. Noe and co-workers have demonstrated that in a high magnetic field perpendicular to a QW, incoherent electron-hole pairs excited by an ultrafast optical pulse spontaneously form a macroscopic dipole and cooperatively interact with each other to emit an intense SF burst~\cite{NoeEtAl2012NP,NoeEtAl2013FP,KimEtAl2013PRB,KimEtAl2013SR,CongEtAl2016JOSABJa}. The high $B$ increased the dipole moment as well as the density of states and suppressed scattering~\cite{BelyaninEtAl1991SSC,BelyaninEtAl1997QSO}.

\begin{figure}[h]
    \centering
    \includegraphics[width=0.48\textwidth]{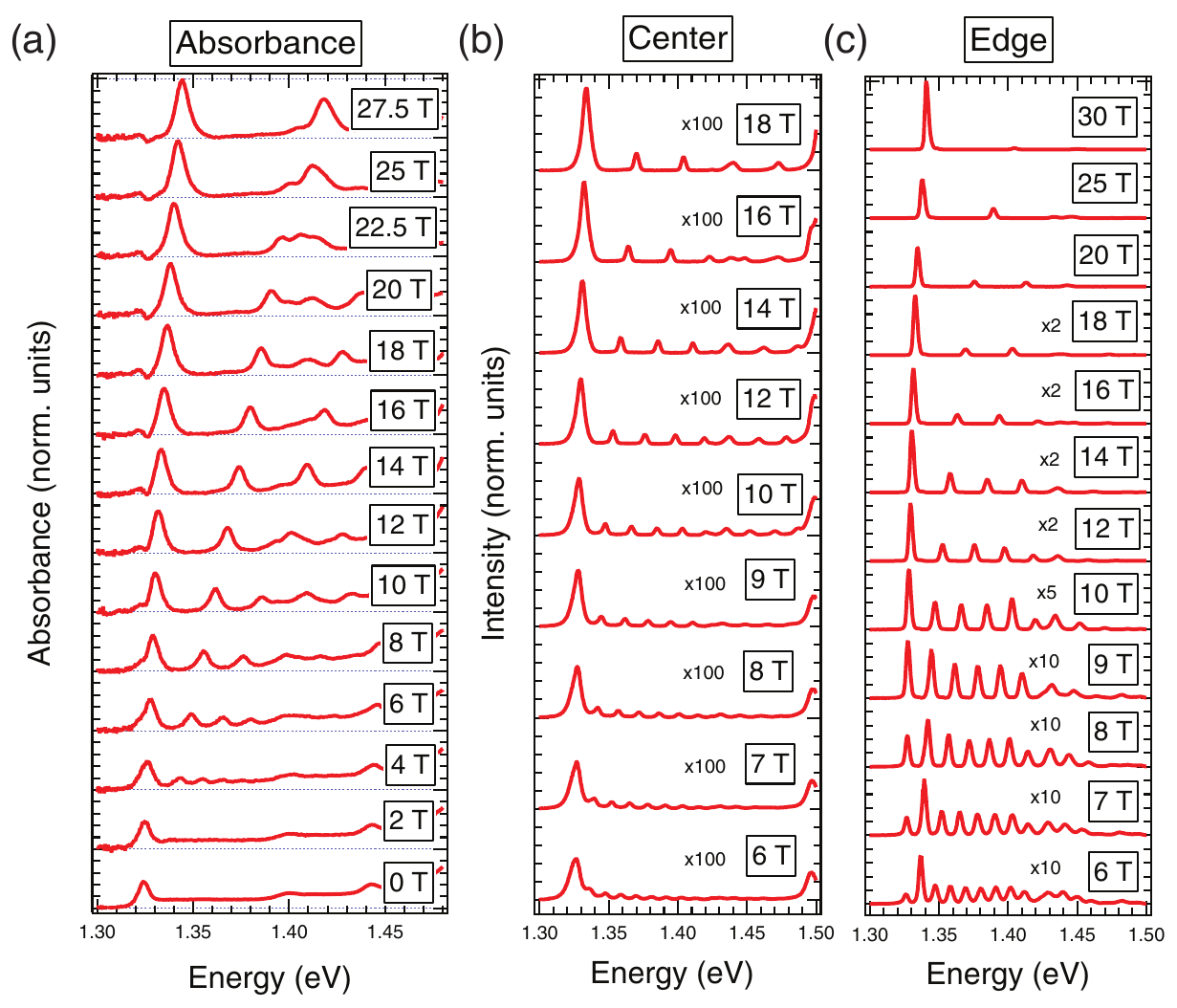}
    \caption{(a)~Absorbance spectra for In$_{0.2}$Ga$_{0.8}$As QW taken at various magnetic fields up to 27.5\,T at 12.5\,K. The traces are vertically offset. Time-integrated magneto-PL spectra for both (b)~center- and (c)~edge-emission at 13\,K. Adapted from Ref.~\citenum{NoeEtAl2013RSI}.}
    \label{fig:SFInGaAs}
\end{figure}

Figure~\ref{fig:SFInGaAs}a displays absorbance spectra for an In$_{0.2}$Ga$_{0.8}$As QW sample at various magnetic fields, taken with the RAMBO system. The LED was used as the light source for the measurements. At 0\,T, the peaks at 1.325\,eV, 1.4\,eV, and 1.442\,eV are attributed to E$_1$H$_1$, E$_1$L$_1$, and E$_2$H$_2$ transitions, respectively~\cite{NoeEtAl2013RSI}. The quasi-2D density of states are quantized into multiple discrete Landau levels when an external magnetic field is applied perpendicular to the QW. The peak frequencies in the absorption spectra increase with increasing magnetic field as the energy separation between LL becomes larger at higher fields.

\begin{figure}[h]
    \centering
    \includegraphics[width=0.45\textwidth]{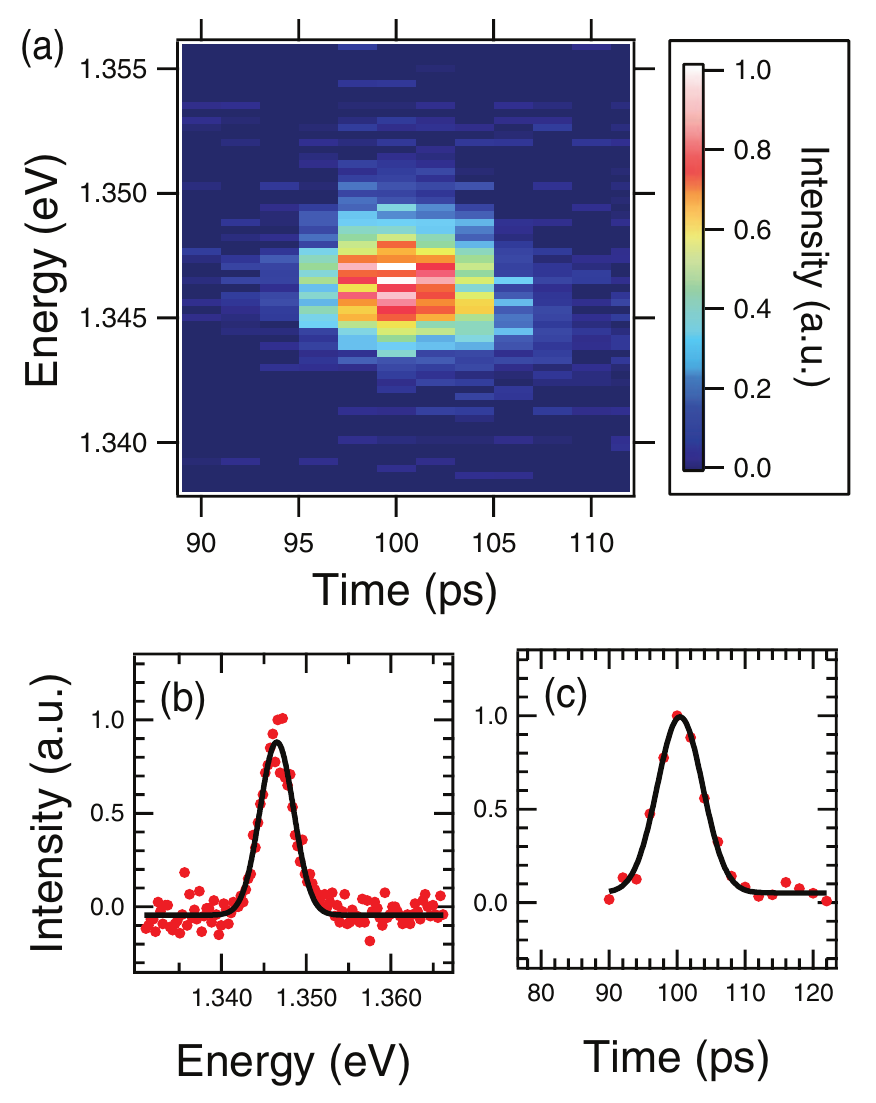}
    \caption{(a)~Time-resolved PL map of emission from the 11 LL transition at 10\,T and at 19\,K for the edge-collected emission. (b)~Spectral and (c)~temporal profile of the superfluorescent burst of radiation. The red points denote the experimental data, while solid lines are fits to the data. Adapted from Ref.~\citenum{NoeEtAl2013RSI}.}
    \label{fig:SFInGaAs2}
\end{figure}

Figures~\ref{fig:SFInGaAs}b and \ref{fig:SFInGaAs}c display time-intagrated PL spectra for center- and edge-collected emission averaged over four single-shot measurements. For the edge-emission measurements, a right-angle microprism was attached to the edge of the sample to redirect the in-plane emission to escape the cryostat~\cite{NoeEtAl2012NP,NoeEtAl2013RSI}. The peaks in the PL spectra correspond to the LL transitions. The peaks shift higher in frequency with increasing $B$, which are consistent with the absorption spectra. In the center-emission spectra, the emission strength of all transitions increase steadily with increasing $B$. By contrast, the edge-collected emission strength of the 00 LL transition with the lowest peak energy increases dramatically from 6\,T to 30\,T. The drastic increase in the edge-emission as compared to the center-emission indicates that the observed edge-emission is SF.

Figure~\ref{fig:SFInGaAs2} shows the edge-collected emission intensity of the time-resolved PL measurement as a function of time and energy at 19\,K and at 10\,T. The pulse duration and spectral width of the SF pulse are determined to be $\sim$10\,ps and $\sim$5\,meV, respectively. The temporal resolution of the setup was limited by the Kerr medium, which is $\sim$1\,ps for toluene~\cite{ChenEtAl2012APL}.

\subsection{Plasmons}
THz spectroscopy is sensitive to free electron dynamics, especially through collective excitation of an electron gas, or plasmons.  In the presence of a magnetic field, a plasmon excitation becomes a magnetoplasmon, whose frequency is given by the cyclotron frequency, $\omega_\text{c} = eB/m^*$, where $m^*$ is the effective mass of the electrons, in the case of a free electron gas with no confinement. In addition, polarization-dependent THz spectroscopy allows one to determine both the diagonal and off-diagonal elements of the conductivity tensor through Faraday and Kerr rotations~\cite{JenkinsEtAl2010PRB,LaForgeEtAl2010PRB,HancockEtAl2011PRL,ValdesAguilarEtAl2012PRL,WuEtAl2015PRL,DziomEtAl2017NC,ChengEtAl2019PRL,LiEtAl2019PRB}. 

\subsubsection{Cyclotron resonance in photoexcited Si}
The RAMBO system has been used for measuring the cyclotron resonance of photoexcited carriers in a single crystal of intrinsic Si at high magnetic fields~\cite{NoeEtAl2016OEO}. The Si sample was excited by near-infrared pulses and then probed by THz pulses after a delay of 100\,ps. The free carriers excited by the pump beam move in circular orbits due to the external magnetic field at the cyclotron frequency, $\omega_\text{c}$, which lies within the THz frequency range in the magnetic field range of RAMBO. 

\begin{figure}[h]
    \centering
    \includegraphics[width=0.45\textwidth]{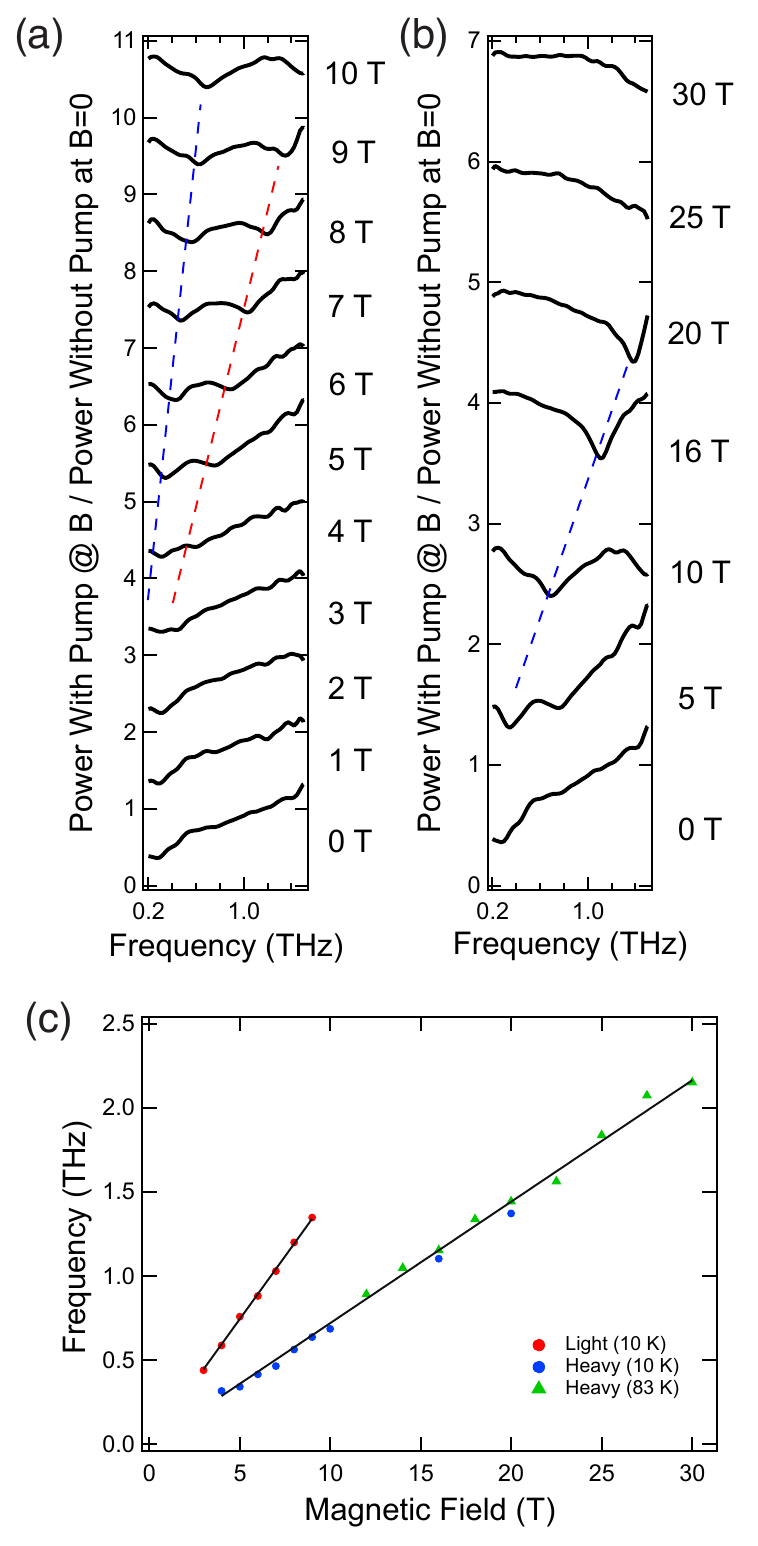}
    \caption{The relative THz transmission of intrinsic Si at 100\,ps after optical excitation at various magnetic fields up to (a)~10\,T and (b)~30\,T, respectively. The data was taken at 10\,K with LiNbO$_3$ generation. (c)~Extracted cyclotron frequencies versus magnetic field. The data taken at 83\,K used ZnTe generation. The solid lines are linear fits to the data. Adapted from Ref.~\citenum{NoeEtAl2016OEO}.}
    \label{fig:SiCR}
\end{figure}

Figures~\ref{fig:SiCR}a and \ref{fig:SiCR}b show the relative THz transmission of the Si sample as a function of magnetic field up to 30\,T at 10\,K with LiNbO$_3$ generation. The dips in the transmittance spectra correspond to the cyclotron resonance. The dip frequency is extracted and plotted in Fig.~\ref{fig:SiCR}c. The measurements taken at 10\,K and 83\,K used, respectively, LiNbO$_3$ and ZnTe crystals to generate THz pulses. From the resonance frequencies, together with the relation $\omega_\text{c}=eB/m^*$, two effective masses, $m^*=0.19m_0$ and $m^*=0.39m_0$, were obtained, corresponding to the light holes and heavy holes, respectively, in Si.

\subsubsection{Faraday and Kerr effects in Bi$_{1-x}$Sb$_{x}$}
The RAMBO system, together with a 10-T superconducting magnet system~\cite{WangEtAl2007OLOa,WangEtAl2010NP,ArikawaEtAl2011PRBa,ArikawaEtAl2012OEOa,ZhangEtAl2014PRLb,ZhangEtAl2016NPa,LiEtAl2018NPb,LiEtAl2018S}, have been used to study the semimetal-to-topological-insulator transition in Bi$_{1-x}$Sb$_{x}$ films. The optical conductivity, polarization rotation angle, $\theta(\omega)$, and ellipticity change, $\eta(\omega)$, of the Faraday and Kerr rotations were extracted from the measurements.

\begin{figure}[h]
    \centering
    \includegraphics[width=0.48\textwidth]{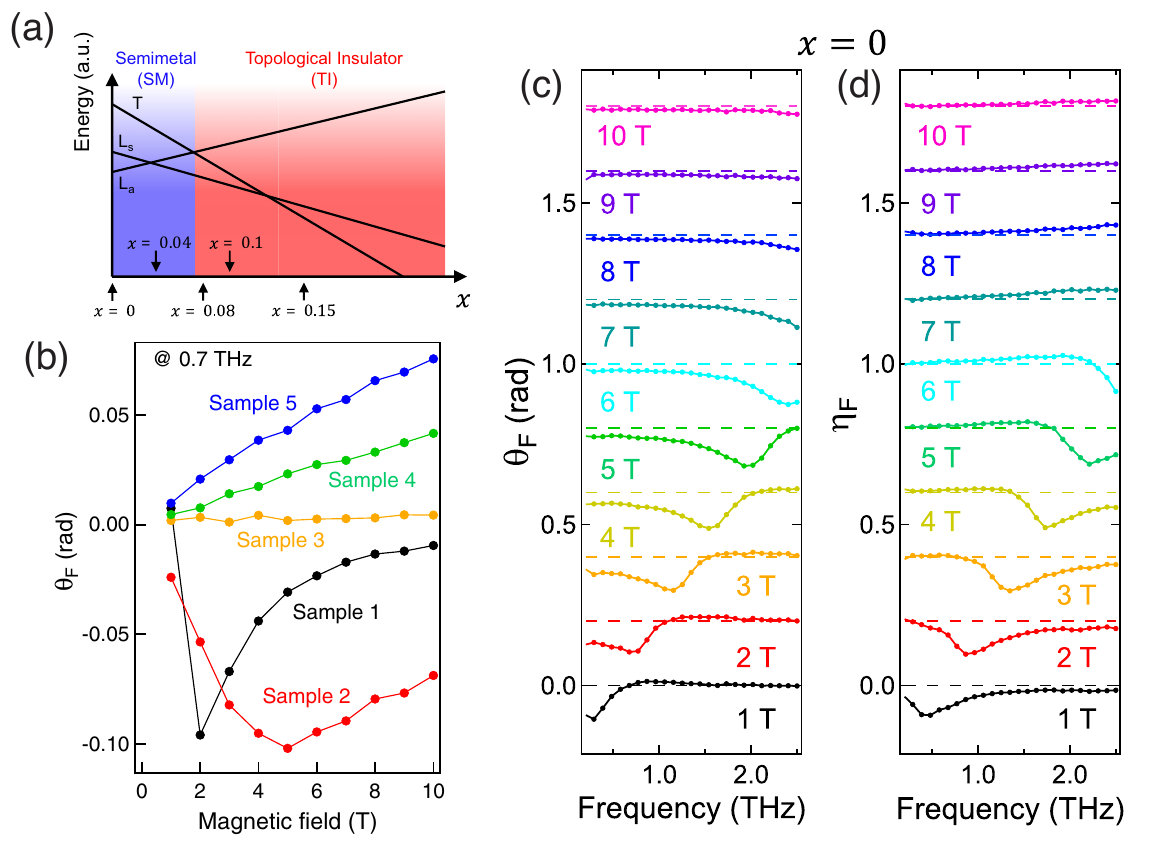}
    \caption{Magneto-optical response of Bi$_{1-x}$Sb$_{x}$ films. (a)~Schematic phase diagram of Bi$_{1-x}$Sb$_{x}$. The five arrows pointing to the horizontal axis mark the doping concentration of the five samples in (b): sample 1\,($x=0)$, sample 2\,($x=0.04$), sample 3\,($x=0.08$), sample 4\,($x=0.1$) and sample 5\,($x=0.15$). The solid lines represent the energies of different band edges at high symmetry points ($T$ and $L$ points). (b)~$\theta_\text{F}$ at 0.7\,THz and at $T=2$\,K for all samples. (c)~$\theta_\text{F}$ and (d)~$\eta_\text{F}$ spectra at various magnetic fields at 2\,K. Adapted from Ref.~\citenum{LiEtAl2019PRB}.}
    \label{fig:BiSb1}
\end{figure}

Figure~\ref{fig:BiSb1}a shows the phase diagram of Bi$_{1-x}$Sb$_{x}$ alloys as a function of $x$, where the transition happens at $x=0.07$~\cite{HsiehEtAl2008N,HsiehEtAl2009N,FuKane2007PRB,TeoEtAl2008PRB,ZhangEtAl2009PRB}. The Bi$_{1-x}$Sb$_{x}$ films were grown on silicon substrates by molecular beam epitaxy~\cite{HiraharaEtAl2010PRB,KatayamaEtAl2018PRB}. The magnetic field dependence of $\theta_\text{F}$ at 0.7\,THz for Bi$_{1-x}$Sb$_{x}$ films with different $x$ exhibits a trend that is consistent with the phase diagram. For semimetallic films (sample 1-2), $\theta_\text{F}$ exhibits a dip below zero within 10\,T. By contrast, $\theta_\text{F}$ becomes positive and increases with increasing $B$ for topological insulating films (sample 3-5). No dip is observed and the change of $\theta_\text{F}$ traces with increasing $x$ is monotonic. 

Figures~\ref{fig:BiSb1}c and \ref{fig:BiSb1}d show $\theta_\text{F}$ and $\eta_\text{F}$ spectra for sample 1\,($x=0$) at various magnetic fields. A resonance feature that shifts to higher frequencies with increasing $B$ is observed in both spectra. Based on a bulk band model with realistic band parameters, the authors concluded that the feature is mainly caused by bulk hole cyclotron resonance~\cite{deVisserEtAl2016PRL,ZhuEtAl2011PRB,LiEtAl2019PRB}.

\begin{figure}[h]
    \centering
    \includegraphics[width=0.45\textwidth]{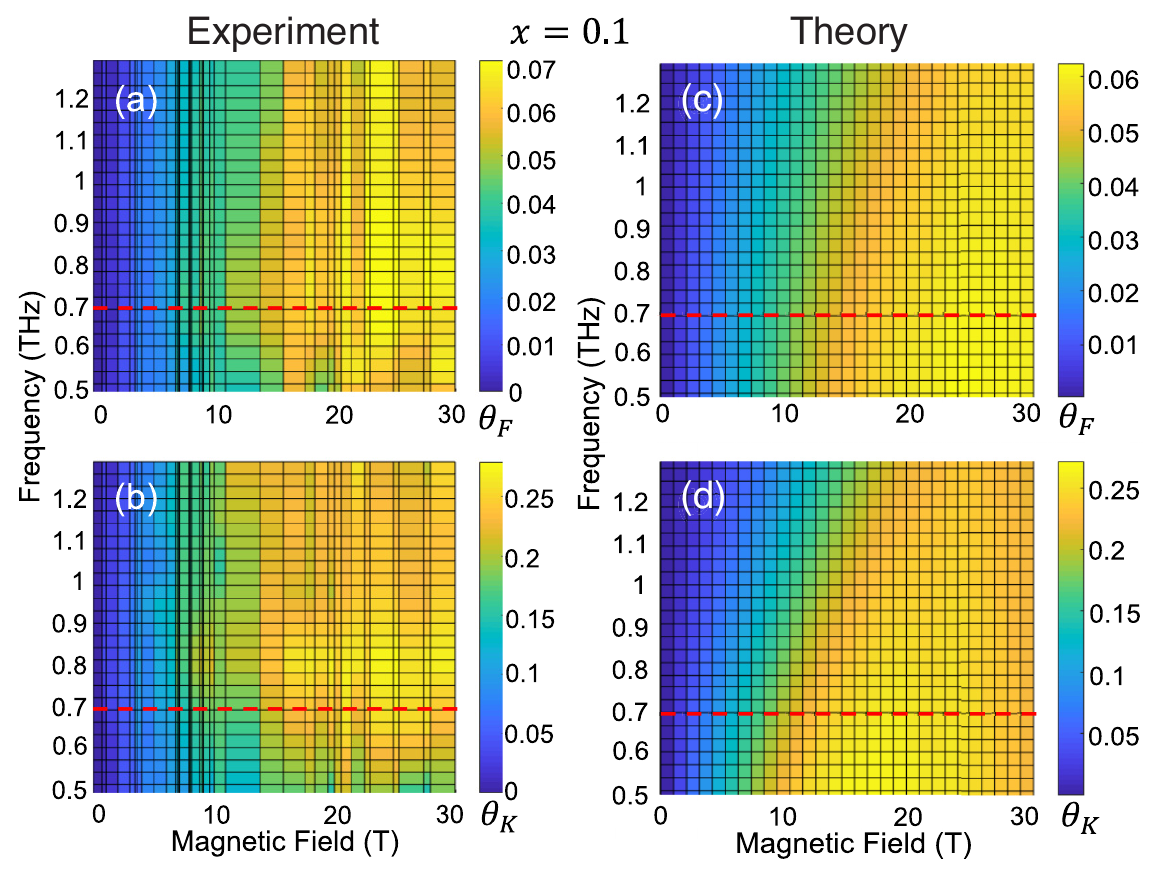}
    \caption{Experimental result of (a)~$\theta_\text{F}$ and (b)~$\theta_\text{K}$ versus frequency and magnetic field for sample 4. Theoretical calculation of (c)~$\theta_\text{F}$ and (d)~$\theta_\text{K}$ spectra at fields between 0 and 30\,T. Adapted from Ref.~\citenum{LiEtAl2019PRB}.}
    \label{fig:BiSb2}
\end{figure}

Figures~\ref{fig:BiSb2}a and \ref{fig:BiSb2}b show, respectively, $\theta_\text{F}(\omega,B)$ and $\theta_\text{K}(\omega,B)$ of the Bi$_{0.9}$Sb$_{0.1}$ film versus frequency and $B$ up to 30\,T at a temperature of 21\,K. At a fixed $B$, both $\theta_\text{F}$ and $\theta_\text{K}$ spectra are featureless. Nevertheless, their values increase with increasing $B$ and finally saturate when $B$ is above 15\,T. 
In contrast to the semimetallic film, a surface band model was used to analyze the experimental data~\cite{BeniaEtAl2015PRB,ZhangEtAl2009PRB,LiEtAl2019PRB}. By summing up the contributions from the $\Bar{\Gamma}$-point electron pocket, the $\Bar{M}$-point electron pocket, and the hole pocket, 2D color maps of $\theta_\text{F}$ and $\theta_\text{K}$ were constructed and shown as Figs.~\ref{fig:BiSb2}c and \ref{fig:BiSb2}d. The theoretical calculations are in agreement with the experimental results as they reproduce the saturation behavior of Faraday and Kerr rotations at high $B$. These results suggest that the combined effort of the THz magnetospectroscopy measurements and the detailed theoretical analysis can be employed to study surface and bulk carrier contributions to the optical conductivity spectra of topological materials.

\subsection{Magnons}
\subsubsection{Ultrastrong magnon-magnon coupling in YFeO$_3$}

\begin{figure}
    \centering
    \includegraphics[width=0.45\textwidth]{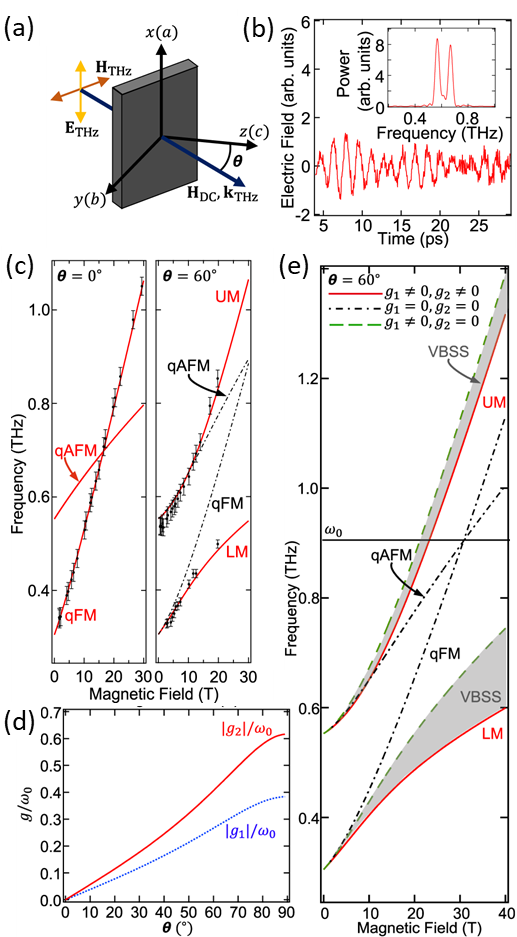}
    \caption{(a)~Schematic of THz magnetospectroscopy studies of YFeO$_3$ in a tilted magnetic field. H$_{\mathrm{DC}}$ was applied in the $b$-$c$ plane at angle $\theta$ with respect to the $c$-axis, with $k_\mathrm{THz}\parallel H_\mathrm{DC}$ and $H_\mathrm{THz}$ polarized in the $b$-$c$ plane. (b)~Transmitted THz waveform for $\theta=20^\circ$ at 12.60~T displaying beating in the time-domain and two peaks in the frequency domain corresponding to the simultaneous excitation of both magnon modes in YFeO$_3$. (c)~Experimentally measured magnon frequencies for $\theta=0^\circ, 60^\circ$ versus H$_\mathrm{DC}$ (black dots) with calculated resonance magnon frequencies (solid red lines) and decoupled qFM and qAFM magnon frequencies (black dashed-dotted lines). The UM frequency becomes lower than the qAFM frequency at $\theta=90^\circ$, indicating a dominant VBSS compared to the vacuum Rabi splitting-induced shifts. (d)~Normalized co-rotating ($|g_1|/\omega_0$, blue dotted line) and counter-rotating ($|g_2|/\omega_0$0, red solid line) coupling strengths displaying ultrastrong magnon–magnon coupling and dominance of the counter-rotating terms. $\omega_0$ is the frequency at which the qFM and qAFM modes cross. (e)~Theoretical illustration of the qFM mode, qAFM mode, lower mode (LM), UM, and co-rotating coupled magnon frequencies that are obtained by setting $g_2=0$, for $\theta=60^\circ$. The vacuum Bloch–Siegert shifts (VBSSs) are highlighted by the shaded area. Adapted from Ref.~\citenum{MakiharaEtAl2021NC}}
    \label{fig:takuma}
\end{figure}

Antiferromagnetic materials host spin waves (magnons) with typical frequencies in the THz range. Makihara et al.\ have studied ultrastrong magnon–magnon coupling dominated by antiresonant interactions in YFeO$_3$~\cite{MakiharaEtAl2021NC}, a canted antiferromagnet that supports quasi-ferromagnetic (qFM) and quasi-antiferromagnetic (qAFM) magnon modes. These magnon modes can be probed with THz radiation if the THz magnetic field component is perpendicular to the spin orientation. Figure~\ref{fig:takuma}a shows a geometry where an external magnetic field, H$_\text{DC}$, is applied at an angle $\theta$ to the $c$-axis. When selection rules are satisfied, both magnon modes can be measured as beating in the time domain and peaks in the frequency domain, as shown in Fig.~\ref{fig:takuma}b.

Figure~\ref{fig:takuma}c shows an example of the evolution of qFM and qAFM modes in an external magnetic field. At $\theta=0$, only the qFM mode is experimentally accessible, while at $\theta=60^\circ$ both magnon modes can be traced as a function of magnetic field. Interestingly, these two modes exhibit anticrossing behaviors, which is an indication of strong coupling/hybridization~\cite{FriskKockumEtAl2019NRP,FornDiazEtAl2019RMP,PeracaEtAl2020SaS}. 

To explain the hybridization between the qFM and qAFM modes, the authors developed a microscopic model, which includes interactions between spins in the two sublattices, their symmetric exchange and antisymmetric exchange, the single-ion anisotropies, and the Zeeman interaction. The system was effectively described by two coupling strengths $g_1$ (co-rotating coupling strength) and $g_2$ (counter-rotating coupling strength). Figure~\ref{fig:takuma}(d) shows normalized coupling strengths as a function of angle $\theta$, where $g_2>g_1$. This led to the observation of a dominant vacuum Bloch-Siegert shift (VBSS) for the upper mode (UM), which is unique to the anisotropic Hopfield Hamiltonian. Figure~\ref{fig:takuma}e summarizes results of numerical calculations demonstrating VBSSs at $\theta=60^\circ$. In addition, quantum fluctuation suppression of up to 5.9\,dB was theoretically showed, indicating two-mode squeezed vacuum magnonic ground state in this system~\cite{MakiharaEtAl2021NC}.

\subsection{Phonons}

\begin{figure}[h]
    \centering
    \includegraphics[width=0.4\textwidth]{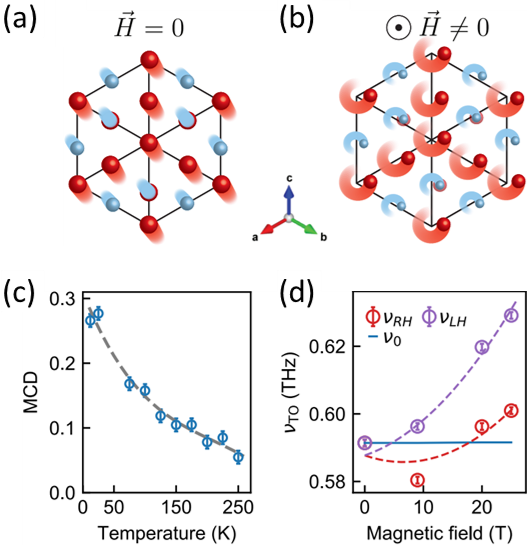}
    \caption{Schematic of the PbTe crystal structure with real-space TO lattice displacements (a)~without and (b)~with a magnetic field applied perpendicular to the lattice plane. Blue (red) spheres represent Te (Pb) ions. (c)~Magnetic circular dichroism at 9\,T as a function of temperature. The dashed line is a guide to the eye. (d)~Magnetic field induced frequency shift of the TO phonon. The dashed lines are fits to the data. Adapted from Ref.~\citenum{BaydinEtAl2021Arxiv}.}
    \label{fig:pbte}
\end{figure}

Usually, phonons are insensitive to magnetic fields. Unlike spectral features related to electrons, which respond to magnetic fields through their orbital and spin magnetic moments, phonon-related features typically remain independent of magnetic fields unless there is strong electron-phonon interaction. Recently, a strong magnetic response of a transverse optical (TO) phonon mode in a thin film of lead telluride (PbTe) has been observed using THz time-domain magnetospectroscopy~\cite{BaydinEtAl2021Arxiv}. PbTe, one of the most widely used thermoelectric materials, is known to have a soft lattice, hosting anharmonic phonons. The displacements of the TO phonon, schematically shown in Fig.~\ref{fig:pbte}a, can become circular under the application of a high magnetic field (see Fig.~\ref{fig:pbte}b). 

The physical picture that emerged via comparison of experimental data and theoretical modelling is that the magnetic field exerted a Lorentz force directly on the ions that form the lattice, inducing ionic cyclotron motion, which created chiral phonons. Figure~\ref{fig:pbte}c shows magnetic circular dichroism of the TO phonon as a function of temperature for a 9\,T applied magnetic field, which means that the response to right- and left-hand
polarized light differs. Figure~\ref{fig:pbte}d summarizes the R- and L-handed TO
phonon frequency as a function of magnetic field, together with fits (dashed lines). This dependence having linear and quadratic terms with respect to the magnetic field indicates a large phononic Zeeman splitting and the existence of a phononic diamagnetic shift. The authors concluded that the magnetic phonon moment comes from TO phonon anharmonicity and morphic effects due to high magnetic fields~\cite{BaydinEtAl2021Arxiv}.

\section{Summary and Outlook}
Overall, since the development of the RAMBO system, we have been able to spectroscopically probe a number of new physical phenomena in high magnetic fields, including superfluorescence in InGaAs quantum wells,\cite{NoeEtAl2013RSI} Faraday and Kerr rotations in a semimetal and topological insulator compound,\cite{LiEtAl2019PRB} ultrastrong magnon-magnon coupling in a rare-earth orthoferrite,\cite{MakiharaEtAl2021NC} and large magnetic moments of phonons in PbTe.\cite{BaydinEtAl2021Arxiv} The easy optical access of RAMBO allows one to optically probe various collective excitations in different energy scales in condensed matter systems by using different light sources. Recently, similar table-top high field magnets have been developed by other groups~\cite{MolterEtAl2010OE,SpencerEtAl2016APL,PostEtAl2021PRB}. The growing interest in the development of such setups is a testament to the promise that more exciting research studies of modern materials under extreme conditions can be performed in university-level laboratories.

While the RAMBO system has been well-established for performing advanced magneto-optical spectroscopy experiments on materials in high magnetic fields, it can be further improved in multiple aspects. First, the mini-coil can be redesigned to support even larger magnetic fields without sacrificing the bore diameter. Second, lower temperatures below 10\,K are highly desirable for some condensed matter samples that show a variety of low-temperature phases. Third, more electrical connections can be introduced for measuring optical and transport properties simultaneously. Finally, combination of RAMBO with even more optical setups covering different spectral ranges will be beneficial for studying interplay between different energy-scale excitations. 

In conclusion, we reviewed the technical specifications of the first-generation RAMBO system and discussed several research advances enabled by RAMBO. As RAMBO-like systems become available to more research groups, we expect to see further new discoveries in materials physics. We believe that, with further development of advanced pulsed magnets, next generation RAMBO systems promise to push the frontiers of materials research in high magnetic fields even further. 



\bibliographystyle{jpsj}
\bibliography{references,rambo,jun}

\begin{wrapfigure}[10]{l}{0.2\textwidth}
    \vspace{-30pt}
    \includegraphics[width=0.2\textwidth]{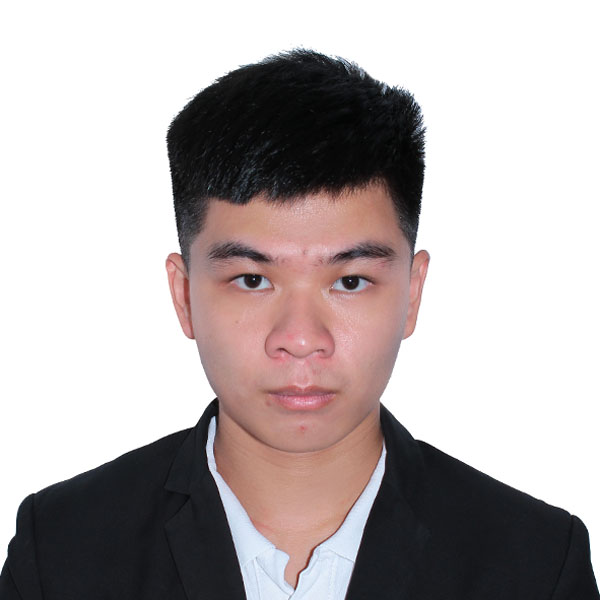}
\end{wrapfigure}
\profile{Fuyang Tay}{is a graduate student in the Applied Physics Graduate Program at Rice University. He received his B.S. degree in Physics from Nanyang Technological University in 2018. His current research interests include ultrastrong light-matter coupling and ultrafast phenomena in condensed matter.\newline \newline}

\begin{wrapfigure}[10]{l}{0.2\textwidth}
    \vspace{-40pt}
    \includegraphics[width=0.2\textwidth]{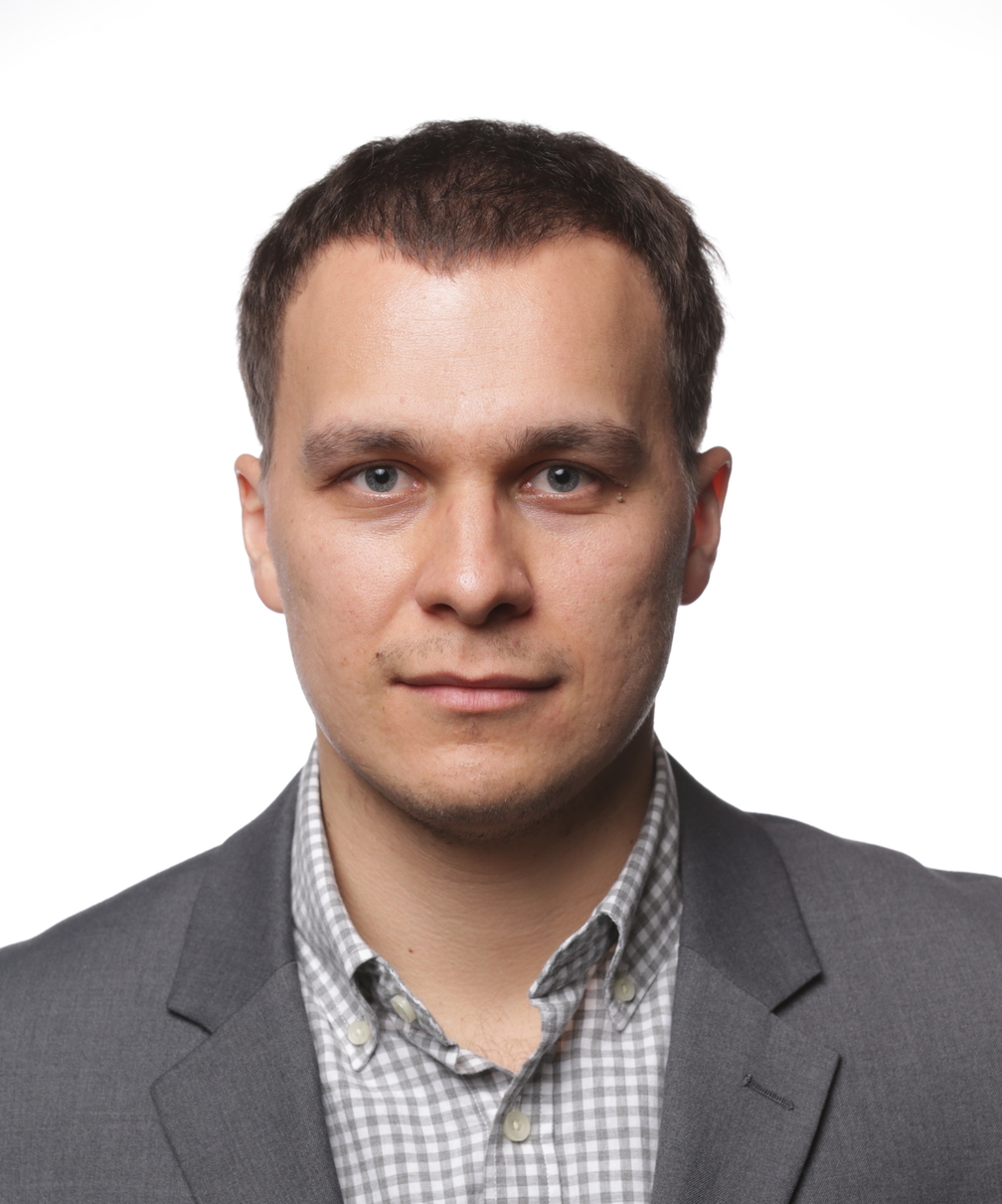}
\end{wrapfigure}
\profile{Andrey Baydin}{is a J. Evans Attwell-Welch Postdoctoral Research Fellow in the Smalley-Curl Institute at Rice University. He obtained his Ph.D. degree in Physics from Vanderbilt University, USA in May 2018. His current research interests include ultrafast spectroscopy of quantum materials and light-matter interaction in the ultrastrong coupling regime. \newline \newline}

\newpage

\begin{wrapfigure}[10]{l}{0.2\textwidth}
    \vspace{-5pt}
    \includegraphics[width=0.2\textwidth]{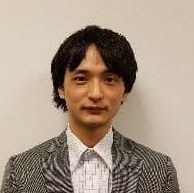}
\end{wrapfigure}
\profile{Fumiya Katsutani}{received his Bachelor’s degree in electrical engineering from National Institution for Academic Degrees and University Evaluation in Japan in 2013 and his Master’s degree in electrical engineering from Osaka University in 2014. He obtained his Ph.D. degree in Electrical and Computer Engineering from Rice University in 2020. Since 2020, he has been an optical engineer \& application scientist at Shimadzu Corporation.\newline \newline \newline \newline \newline}

\begin{wrapfigure}[12]{l}{0.2\textwidth}
    \vspace{-30pt}
    \includegraphics[width=0.2\textwidth]{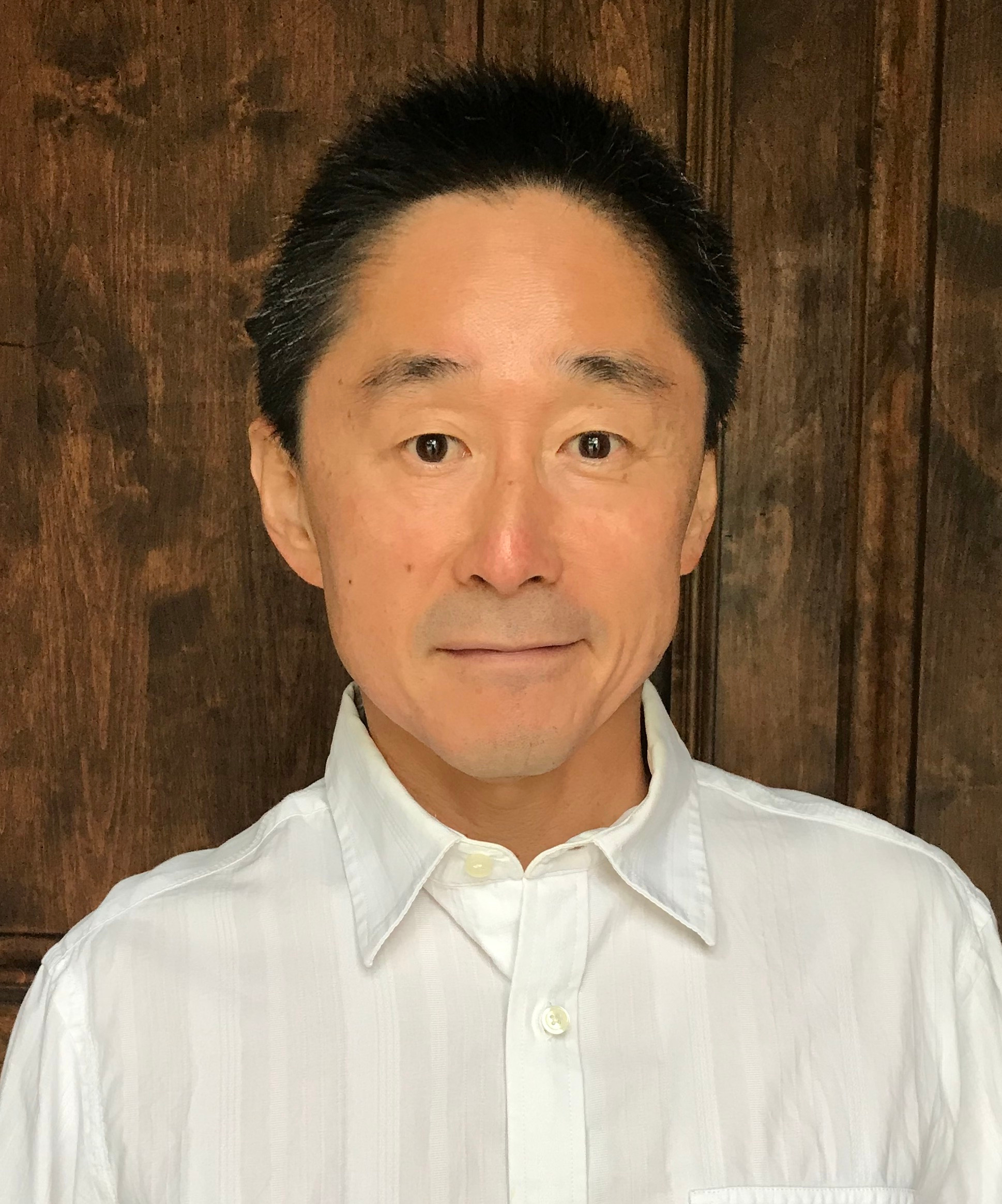}
\end{wrapfigure}
\profile{Junichiro Kono}{is Karl F.\ Hasselmann Chair in Engineering, Professor in the Departments of Electrical \& Computer Engineering, Physics \& Astronomy, and Materials Science \& Nanoengineering, and Chair of Applied Physics at Rice University. He received his B.S.\ and M.S.\ degrees in applied physics from the University of Tokyo in 1990 and 1992, respectively, and completed his Ph.D.\ in physics at the State University of New York at Buffalo in 1995. He was a postdoctoral research associate at the University of California Santa Barbara from 1995-1997, and the W.\ W.\ Hansen Experimental Physics Laboratory Fellow in the Department of Physics at Stanford University from 1997-2000. His current research interests include quantum optics in condensed matter, ultrastrong light-matter coupling, and terahertz science and technology.}

\end{document}